\documentclass{article}

\usepackage{amssymb,amsfonts,amsmath}
\usepackage{cite,enumerate,float}
\usepackage{color}
\usepackage{tikz}
\usetikzlibrary{arrows,snakes,backgrounds}
\usepackage{url}
\usepackage[vcentermath]{youngtab}

\def\be{\begin{eqnarray}}
\def\ee{\end{eqnarray}}
\def\nn{\nonumber}

\def\p{\partial}

\def\wcs{weak compositions\ }
\def\Id{{\rm Id}}

\def\NS{\mathfrak{P}}

\definecolor{red}{rgb}{1,0,0}
\definecolor{orange}{rgb}{1,0.5,0}
\definecolor{violet}{rgb}{0.7,0,1}

\newcommand*{\Searrow}{\rotatebox[origin=c]{-45}{\(\xrightarrow{\hspace*{.3cm}}\)}}

\newcommand*{\Swarrow}{\rotatebox[origin=c]{225}{\(\xrightarrow{\hspace*{.3cm}}\)}}

\def\triad{{\footnotesize\fbox{$\begin{array}{rcccl}
&&\!\!\!\!\!\!\!\!\!\!\!\!\!{\bf NS^v}&&\\
&\!\!\!\!\!\!\!\!\!\!\Swarrow&&\!\!\!\!\!\!\!\!\!\!\!\!\!\!\!\!\!\!\!\!\Searrow& \\
{\bf M}&&\!\!\!\!\!\!\!\!\!\!\!\!\!\!\!\longleftarrow&&\!\!\!\!\!\!\!\!\!\!\!\!\!{\bf BA^v}
\end{array}$}}}

\def\Triad{{\footnotesize\fbox{$\begin{array}{rcccl}
&&\!\!\!\!\!\!\!\!\!\!{\bf \mathfrak{E}^{(w,v)}}&&\\
&\!\!\!\!\!\!\!\!\!\!\Swarrow&&\!\!\!\!\!\!\!\!\!\!\!\!\!\!\!\!\!\!\!\!\Searrow& \\
{\bf E}&&\!\!\!\!\!\!\!\!\!\!\!\!\!\!\!\longleftarrow&&\!\!\!\!\!\!\!\!\!\!\!\!\!{\bf \Psi^{(w,v)}}
\end{array}$}}}

\hoffset= - 1.0in         

\begin{document}

\title{\vspace{1.5cm}\bf Non-symmetric triads and Baker-Akhiezer functions
}

\author{
A. Mironov$^{b,c,d,}$\footnote{mironov@lpi.ru,mironov@itep.ru},
A. Morozov$^{a,c,d,}$\footnote{morozov@itep.ru},
A. Popolitov$^{a,c,d,}$\footnote{popolit@gmail.com}
}

\date{ }

\maketitle

\vspace{-6cm}

\begin{center}
FIAN/TD-16/26  \hfill {\bf to the memory of}\\
ITEP/TH-23/26   \hfill {\bf Sergei Kharchev}\\
IITP/TH-21/26  \hfill \phantom{.}\\
MIPT/TH-19/26 \hfill \phantom{.}
\end{center}

\vspace{4.5cm}

\begin{center}
$^a$ {\small {\it MIPT, Dolgoprudny, 141701, Russia}}\\
$^b$ {\small {\it Lebedev Physics Institute, Moscow 119991, Russia}}\\
$^c$ {\small {\it NRC ``Kurchatov Institute", 123182, Moscow, Russia}}\\
$^d$ {\small {\it Institute for Information Transmission Problems, Moscow 127994, Russia}}
\end{center}

\vspace{.1cm}

\begin{abstract}
The {\bf symmetric} Macdonald polynomial at peculiar values of parameter $t=q^{-m}$, $m\in\mathbb{Z}_{\ge 0}$ is naturally split into non-symmetric parts, which are the (quasi)polynomial Baker-Akhiezer (BA) functions. One may think this is due to symmetricity, and one just picks up this way non-symmetric parts already containing all the information. However, we demonstrate that, in the case of {\bf non-symmetric} Macdonald polynomials, it still works, though each single BA function splits into $N!$ distinct (quasi)polynomial BA functions. The sum of these functions gives rise to the universal solution of the eigenstate problem for the Cherednik Hamiltonians.
Extending to arbitrary values of $t$ is also immediate giving rise to counterparts of the Noumi-Shiraishi power series.
Altogether, this power series  and its reductions to non-symmetric Macdonald polynomials and to BA functions form a non-symmetric triad.
There are $N!$ different branches of the non-symmetric triad, each branch being split into $N!$ distinct triads, and of these $(N!)^2$ triads $N!(N-1)!$ are independent. We describe in detail the simplest $N=2$ case.
\end{abstract}

\bigskip

\section{Introduction}

The problem of constructing eigenfunctions of Ruijsenaars trigonometric integrable systems has been studied for decades, and has led to discovery of many new and important directions. The simplest eigenfunctions are symmetric Macdonald polynomials $M_\mu(\vec x)$ \cite{Macbook}, which are symmetric polynomials of $N$ variables $x_1,\ldots,x_N$, and are common eigenfunctions of
$N$ commuting Ruijsenaars-Schneider Hamiltonians. They form a discrete spectrum of solutions to the eigenstate problem parameterized by Young diagrams (partitions) $\mu$. However, there are many more solutions, some of them can be realized via specific integral representations \cite{Khar0,Khar1,Khar2,Khar3,Khar4}. Another approach is due to M. Noumi and J. Shiraishi \cite{NS}, who proposed a power series that gives an eigenfunction of the Ruijsenaars-Schneider Hamiltonians with arbitrary eigenvalues providing a continuous spectrum of solutions to the eigenstate problem. Their approach was basing on an analytic continuation of the branching rules of the symmetric Macdonald polynomials \cite{Macbook} to complex values of lines $\mu_i$'s of Young diagrams that parameterize the Macdonald polynomials. This continuation was immediate, since the branching rules have a factorized form of ratios of $q$-Pochhammer symbols. As was realized later \cite{MMP3} (see also \cite{dFK2}), the Noumi-Shiraishi (NS) power series has not only the polynomial reduction to the symmetric Macdonald polynomials at specific values of $\mu_i$'s, but also another (quasi)polynomial reduction, still at {\bf arbitrary} $\mu$ but at specific values of parameter of the system $t=q^{-m}$, $m\in\mathbb{Z}_{\ge 0}$. The polynomials emerging are nothing but the multivariable Baker-Akhiezer function introduced earlier by O. Chalykh \cite{Cha}.

\paragraph{Symmetric triad.} By construction, the NS series and the both polynomial reductions are eigenfunctions of the Ruijsenaars-Schneider Hamiltonians, and they create {\bf symmetric triad}\footnote{The name refers to existence of the symmetric polynomial reduction, to the symmetric Macdonald polynomials, in variance with the non-symmetric triad below, which does not admit symmetric polynomial reductions.}  \triad. Note that the two polynomial reduction are connected with each other at $t=q^{-m}$ and specific $\mu$ associated with a Young diagram: the sum of the BA functions over permutations of coordinates gives rise to the Macdonald polynomial. In the meanwhile, the degrees of the Macdonald polynomials are restricted by $\mu_i$'s, while those of the BA function by $m$.
As a result, at large enough $\mu_i$'s as compared with $m$, the Macdonald polynomial clearly decays into well separated pieces associated with the BA functions.

Note that the NS function is not symmetric, and one can write down different power series depending on the choice of order of $x_i$'s. Each choice is related with the concrete Weyl chamber. Let us choose a ``canonical" order to be $x_1>x_2>\ldots >x_N$, i.e. the corresponding power series is a series in rations ${x_i\over x_j}$ with $i>j$. Then, any other choice is generated by a permutation $v$, which we use as a superscript of the NS function (and similarly for the BA function): 
\be\label{v} 
\NS^v(\vec x;\vec y;q,t)=\NS^{\rm Id}(v(\vec x);\vec y;q,t)
\ee

As an illustration, consider the 2-particle case. In this case, the NS power series is \cite{NS}
\be\label{NS}
\NS^{{\rm Id}}(x_1,x_2;y_1,y_2;q,t)=x_1^{\lambda_1}x_2^{\lambda_2}\cdot t^{z_1-z_2\over 2}\cdot
\sum_{k=0}\left({x_2\over x_1}\right)^k\prod_{j=1}^{k}{(1-tq^{j-1})
\over (1-q^j)}{(1-t^{-1}q^{1-j}y_1y_2^{-1})
\over (1-q^{-j}y_1y_2^{-1})}
\ee  
Here we use the variables $y_i:=q^{\lambda_i}=q^{\mu_i-N+i}=q^{\mu_i+i-2}$, since, in these variables, the power series does not change under the permutation of $\vec x$ and $\vec y$ \cite{NS}. Throughout the paper, we also use the notation $x_i=q^{z_i}$. 

At integer $\mu_1\ge\mu_2\ge 0$, the sum is restricted by $\mu_1-\mu_2$, and $\NS^{{\rm Id}}(x_1,x_2;y_1,y_2;q,t)$ becomes the symmetric Macdonald polynomial $M_\mu(\vec x;q,t)$. Similarly, at $t^{-m}$, $m\in\mathbb{Z}_{\ge 0}$, the sum is restricted by $m$, and it becomes a (quasi)polynomial BA function $\Psi_m(\vec x;\vec y;q)$, i.e. a polynomial of degree $m$ multiplied by a specific monomial factor.
All these quantities are eigenfunctions of the Ruijsenaars-Schneider Hamiltonians \cite{RS1,RS2} (= the Macdonald operator \cite{MRS1,MRS2,MRS3}), the first of them being
\be\label{RS}
H_1^{RS}=\sum_{i=1}^N\prod_{j\ne i}{tx_i-x_j\over x_i-x_j}q^{\hat D_i}\nn
\ee
where $\hat D_i:=x_i{\p\over\p x_i}$. Note that one can equally well consider the power series $\NS^{\rm Id}(x_2,x_1;y_1,y_2;q,t)$, which is, of course, a distinct power series being power series of the distinct ratio. However, it is still an eigenfunction of the same Ruijsenaars-Schneider Hamiltonians with the same eigenvalue, since the Hamiltonians are symmetric in $x_i$'s. The coefficients of the both power series are the same, just the variables are permuted. This essentially differs from the non-symmetric case below, where the coefficients of the corresponding power series are also distinct.

At $t=q^{-m}$, the symmetric Macdonald polynomial can be also obtained as a sum of two BA functions with permuted variables $x_1\leftrightarrow x_2$:
\be 
M_{[\mu_1,\mu_2]}(x_1,x_2;q,q^{-m})=\Psi_m(x_1,x_2;y_1,y_2;q)+\Psi_m(x_2,x_1;y_1,y_2;q)\nn
\ee
Since the Macdonald polynomial is a polynomial of degree $\mu_1-\mu_2$, if one chooses $\mu_1-\mu_2>m$, the two BA polynomials are clearly parted, since each of them is a polynomial of degree $m$. For instance, at $m=1$,
\be 
M_{[6,0]}(x_1,x_2;q,q^{-1})=\underbrace{x_1^6-{(q^6-1)\over q(q^4-1)}x_1^5x_2}_{\Psi_1(x_1,x_2;q^5,1;q)}+0\cdot x_1^4x_2^2+0\cdot x_1^3x_2^3+0\cdot x_1^2x_2^4\underbrace{-{(q^6-1)\over q(q^4-1)}x_1x_2^5+x_2^6}_{\Psi_1(x_2,x_1;q^5,1;q)}\nn
\ee

\paragraph{Non-symmetric triad.} Similarly to the trigonometric Ruijsenaars system, another integrable system given by the Cherednik Hamiltonians admits polynomial solutions, the non-symmetric Macdonald polynomials $E_\mu(\vec x)$ \cite{Opd95,Mac96,Che95,MN}. They form a discrete spectrum of solutions to the eigenstate problem parameterized by {\bf weak compositions} $\mu$.

An analytic continuation to complex values of $\mu$ is still also possible, but it is multi-valued \cite{NSM5}. Since the structure of non-symmetric Macdonald polynomials crucially depends on order of the elements of $\mu$, and there are $N!$ different orders, there are $N!$ different continuations, which we call different branches. In fact, there is a larger ambiguity: even for the given branch, there are $N!$ possibilities of continuing to the power series depending on which variables the power series should be based on: it can be a power series, say, of ratios ${x_i\over x_j}$ for all $i<j$ (as in (\ref{NS}). Other possibilities can be enumerated by the permutations of $x_i$ in these ratios, and there are $N!$ permutations. In fact, every such set of ratios is associated with a concrete Weyl chamber. The crucial difference with the symmetric case is that now the power series for different Weyl chambers turn out to be distinct: they are not related by permutations of $x_i$'s as in (\ref{v}), they are essentially different power series.

Hence, the power series in the non-symmetric case is labeled by two elements $w$ and $v$ of the permutation group $S_N$: $\mathfrak{E}^{(w,v)}(\vec x;\vec y;q,t)$. We associate the first element $w$ with labelling the branch, and the second one $v$ with the Weyl chamber.

An essential difference with the symmetric case, where the power series with any order was an eigenfunction of the Ruijsenaars-Schneider Hamiltonians (because of the symmetricity), in the non-symmetric case, the eigenfunction is only given by the sum over all Weyl chambers. This is what should be called {\bf universal solution} \cite{dFK2} in the non-symmetric case: 
\be 
{\cal E}^w(\vec x;\vec y;q,t)=\sum_{v\in S_N}\mathfrak{E}^{(w,v)}(\vec x;\vec y;q,t)\nn
\ee

Thus, there are totally $N!$ branches of power series, and $N!$ possible Weyl chambers for each branch. This gives us totally $(N!)^2$ power series, and each of them admits the two polynomial reductions again: to $N!$ the non-symmetric Macdonald polynomials at non-negative integer values of $\mu_i$'s (since any branch gives rise to the same non-symmetric Macdonald polynomial), and to $(N!)^2$ (quasi)polynomial BA functions. Hence, we come to the notion of {\bf non-symmetric triad} \Triad.

\bigskip

In fact, only $N!(N-1)!$ out of these $(N!)^2$ triads are independent: they are related by formula
\be\label{cyc}
{\cal E}^{v_cw}(x_1,x_2,\ldots,x_N;y_2,\ldots,y_N,qy_1;q,t)=
t^{-w(\rho)_1}y_1x_N{\cal E}^w(qx_{n},x_1,x_2,\ldots,x_{N-1};y_1,\ldots,y_N;q,t)
\ee
where $v_c$ denotes the cyclic permutation $[1,2,\ldots,N]\to[2,\ldots,N,1]$, and this identity is inherited from the well-known formula for the non-symmetric Macdonald polynomials (part of the Knop-Sahi recurrence \cite{KS})
\be
E_{[\mu_2,\ldots,\mu_N,\mu_1+1]}(x_1,x_2,\ldots,x_N)=
q^{-\mu_1}x_NE_{[\mu_1\mu_2,\ldots,\mu_N]}(qx_{n},x_1,x_2,\ldots,x_{N-1})\nn
\ee
Because of (\ref{cyc}), the triads from the same cyclic orbit of $S_N$, i.e. those obtained by cyclic permutations are proportional to each other, and there are exactly $(N-1)!$ cyclic orbits, which gives us the number of independent triads $N!(N-1)!$. In fact, they are not completely independent: they are just related by more involved formulas, see sec.\ref{5.4.2} below.

Again, in the non-symmetric triad, the two polynomial reductions are connected with each other at $t=q^{-m}$ and specific $\mu$ associated with a weak composition: the sum of the BA functions over all Weyl chambers gives rise to the Macdonald polynomial:
\be 
E_\mu(\vec x;q,q^{-m})=\sum_{v\in S_N}\Psi_m^{(w,v)}(\vec x;\vec y;q)\nn
\ee
Notice how summing over all permutations with unit coe4fficients still gives rise to a non-symmetric result.
Here $w$ marks the type of order\footnote{By the type of order for the weak composition $\mu$ of length $N$, we mean such sequence $\{n_i\}$ of the first $N$ natural numbers that, for any $i>j$, if $\mu_i\ge \mu_j$, then $n_i>n_j$, and if $\mu_i< \mu_j$, then $n_i<n_j$.}
of non-negative integers in the weak composition $\mu$.
And again, at large enough $\mu_i$'s as compared with $m$, the non-symmetric Macdonald polynomial clearly decays into well separated pieces associated with the BA functions for different Weyl chambers. 

In the $N=2$ case, all this means that there are two branches of power series $\mathfrak{E}(x_1,x_2;y_1,y_2)$, each of them associated with two possible Weyl chambers (see (\ref{NSE}) and (\ref{rels})), each of these four power series gives rise to the corresponding (quasi)polynomial BA function at $t=q^{-m}$, $m\in\mathbb{Z}_{\ge 0}$. The other polynomial reduction is provided when both $\mu_1$ and $\mu_2$ are non-negative integers, and, in this case, the different Weyl chambers gives rise to the same non-symmetric Macdonald polynomial: $\mathfrak{E}^{([2,1],[1,2])}\Big(x_1,x_2;\lambda_1,\lambda_2\Big)$ and $\mathfrak{E}^{([2,1],[2,1])}\Big(x_1,x_2;\lambda_1,\lambda_2\Big)$ gives the same polynomial $E_{[\mu_1,\mu_2]}(x_1,x_2)$ (\ref{E12}) with $\mu_1\ge\mu_2$, and $\mathfrak{E}^{([1,2],[1,2])}\Big(x_1,x_2;\lambda_1,\lambda_2\Big)$ and $\mathfrak{E}^{([1,2],[2,1])}\Big(x_1,x_2;\lambda_1,\lambda_2\Big)$ gives the same polynomial $E_{[\mu_1,\mu_2]}(x_1,x_2)$ (\ref{E21}) with $\mu_2>\mu_1$. However, $\mathfrak{E}^{([2,1],[1,2])}\Big(x_1,x_2;\lambda_1,\lambda_2\Big)$ is a power series of the ratio ${x_1\over x_2}$, while $\mathfrak{E}^{([2,1],[2,1])}\Big(x_1,x_2;\lambda_1,\lambda_2\Big)$, that of ${x_2\over x_1}$ (and similarly for $\mathfrak{E}^{([1,2],*)}\Big(x_1,x_2;\lambda_1,\lambda_2\Big)$). Note that these two series are {\bf distinct}, while in the symmetric case there was {\bf the same} power series, but just of the inverse ratio.

At the same time, the sum over the Weyl chambers,
\be
{\cal E}^{[2,1]}(x_1,x_2;\lambda_1,\lambda_2)=\mathfrak{E}^{([2,1],[1,2])}\Big(x_1,x_2;\lambda_1,\lambda_2\Big)+
{\mathfrak{E}}^{([2,1],[2,1])}\Big(x_1,x_2;\lambda_1,\lambda_2\Big)\nn\\
{\cal E}^{[1,2]}(x_1,x_2;\lambda_1,\lambda_2)=\mathfrak{E}^{([1,2],[1,2])}\Big(x_1,x_2;\lambda_1,\lambda_2\Big)+
{\mathfrak{E}}^{([1,2],[2,1])}\Big(x_1,x_2;\lambda_1,\lambda_2\Big)\nn
\ee
gives rise to eigenfunctions of the Cherednik Hamiltonians
\be\label{Ch}
C_1={(tx_1-x_2)\over (x_1-x_2)}q^{\hat D_1}+{(1-t)x_2\over (x_1-x_2)}\sigma_1q^{\hat D_1}\nn\\
C_2=q^{\hat D_2}{(x_1-tx_2)\over (x_1-x_2)}-q^{\hat D_2}{(1-t)x_2\over (x_1-x_2)}\sigma_1\nn
\ee
($\sigma_i$ permutes $x_i$ and $x_{i+1}$) with arbitrary eigenvalues (i.e. $\lambda_i$'s are in no way restricted):
\be
C_i\cdot{\cal E}(x_1,x_2;\lambda_1,\lambda_2)=q^{\lambda_i}t^{1\over 2}\ {\cal E}(x_1,x_2;\lambda_1,\lambda_2)\nn
\ee

The goal of this paper is to explain in detail the scheme that we just sketched in the Introduction, and to illustrate it with examples.

The paper is organized as follows. In section 2, we study examples of non-symmetric Macdonald polynomials at $N=2$ and $N=3$ and $t=q^{-1}$, $t=q^{-2}$ in order to see the decay of these polynomials into a few BA functions, and to get a flavour of the generic structures in the theory. Then, in section 3, we go to formal algebraic constructions, and define the standard DAHA framework for the non-symmetric Macdonald polynomials. In section 4, we consider a particular example of $N=2$ in the very detail. At last, in section 5, we describe generic non-symmetric triads at arbitrary $N$. Section 6 contains some concluding remarks.

\paragraph{Notation.} The quantum numbers are defined to be
\be
[n]_q:={q^n-1\over q-1}\nn
\ee

Throughout the paper, we equivalently use for a set of variables both the notation $\{\xi_i\}$ and $\vec \xi$. 

For two sets $\vec x$ and $\vec\mu$, we use the notation $x^\mu:=\prod_i x_i^{\mu_i}$.

We use the notation $x_i=q^{z_i}$. The power series $\mathfrak{E}(\vec x;\vec y)$ and the BA functions $\Psi_m(\vec x;\vec\lambda)$ are usually considered as functions of $y_i=q^{\lambda_i}$, while the non-symmetric Macdonald polynomials $E_\mu(\vec x)$ are functions of $\mu=\{\mu_i\}$, which is a weak composition. The weak composition $\mu$ reordered in such a way that it becomes a Young diagram is denoted $\mu^+$. Relation between $\vec\lambda$ and $\vec\mu$ may depend on the concrete framework, but it is always $\lambda=\mu+$ shift that does not depend on $\lambda$, $\mu$.

In the paper, we keep the notation $m$ for the non-negative integer emerging within the specification $t=q^{-m}$ without special mentioning.

We also use throughout the paper the notation $[N,N-1,\ldots,1]$ to label a set of quantities ordered in accordance $\xi_N\ge\xi_{N-1}\ge\ldots\ge\xi_1$, and similarly for other orders.

\section{Hints on the non-symmetric triad and Baker-Akhiezer functions}

As we already pointed out above, essential facts that allow one to imply the presence of BA function are that the Macdonald polynomial is a sum of the BA functions, and that the BA function is a (quasi)polynomial of degree $m$ in any variable, while the Macdonald polynomial is a polynomial of degrees given by the Young diagram in the symmetric case and by the weak composition in the non-symmetric one. This means that if the elements of the Young diagram/weak composition are large enough as compared with $m$, many coefficients in the polynomial must vanish, and hence it splits into separate parts associated with the BA functions. In this section, we demonstrate how this works, and this will give us a clear hint on what the suitable BA functions are.

\subsection{$N=2$ case}

We start from the simplest case of $N=2$. The non-symmetric Macdonald polynomial in this case is labeled by two non-negative integers $\mu_{1,2}$, and
\be
\left\{
\begin{array}{cc}
E_{[\mu_1,\mu_2]}=(x_1x_2)^{\mu_2}E_{[\mu_1-\mu_2,0]}&\ \ \ \ \ \ \ \ \ \ \hbox{if }\mu_1\ge\mu_2\\
\\
E_{[\mu_1,\mu_2]}=(x_1x_2)^{\mu_1}E_{[0,\mu_2-\mu_1]}&\ \ \ \ \ \ \ \ \ \ \hbox{if }\mu_2>\mu_1
\end{array}
\right.\nn
\ee
This means that it suffices to consider simply $E_{[\mu,0]}$ and $E_{[0,\mu]}$, i.e. that there are two polynomials associated with two possible types of order [2,1] and [1,2] respectively, but each of these polynomials depends only on one $\mu$.

In this case, in order to observe the phenomenon of splitting the polynomial into separate BA functions (at $N=2$, there should be two BA functions, i.e. two pieces), one has to choose $t=q^{-m}$ with $\mu>2m$. Hence, the first interesting examples are
\be
E_{[4,0]} &=& x_1^4 + \frac{q[4]_q(t - 1)}{(q^4t-1) }x_1^3x_2+ \frac{q^2[4]_q[3]_q(t - 1)(qt-1)}{[2]_q(q^3t-1)(q^4t-1)}x_1^2x_2^2
+ \frac{q^3[4]_q(t-1)(qt-1)}{(q^3t-1)(q^4t-1)}x_1x_2^3
+ \frac{q^4(t - 1)}{q^4t-1}x_2^4\nn\\
E_{[0,5]} &=& \frac{t-1}{q^4t-1} x_1^4x_2
+\frac{[4]_q(t-1)(qt-1)}{(q^3t-1)(q^4t-1)}x_1^3x_2^2
+ \frac{[4]_q[3]_q(t-1)(qt-1)}{[2]_q(q^3t-1)(q^4t-1)}x_1^2x_2^3 + \frac{ [4]_q(t - 1) }{(q^4t-1) }x_1x_2^4+x_2^5\nn\\
E_{[5,0]} &=& x_1^5
+ \frac{q[5]_q(t - 1)}{(q^5t-1) }x_1^4x_2
 + \frac{q^2[5]_q[4]_q (t - 1)(qt-1)}{[2]_q(q^4t-1)(q^5t-1)}x_1^3x_2^2
+ \frac{q^3[5]_q[4]_q(q^2t-1)(qt-1)(t-1)}{[2]_q(q^3t-1)(q^4t-1)(q^5t-1)}x_1^2x_2^3+ \nn\\
&+& \frac{q^4[5]_q(t-1)(qt-1)}{(q^4t-1)(q^5t-1)}x_1x_2^4+\frac{q^5(t - 1)}{q^5t-1}x_2^5
\nn \\
E_{[0,6]}&=&\frac{t-1}{q^5t-1} x_1^5x_2+{[5]_q(t-1)(qt-1)\over(q^5t-1)(q^4t-1)}x_1^4x_2^2+
{[4]_q[5]_q(t-1)(qt-1)(q^2t-1)\over [2]_q(q^5t-1)(q^4t-1)(q^3t-1)}x_1^3x_2^3+\nn\\
&+&{[4]_q[5]_q(t-1)(qt-1)\over [2]_q(q^5t-1)(q^4t-1)}x_1^2x_2^4+{[5]_q(t-1)\over(q^5t-1)}x_1x_2^5+x_2^6\nn
\ee
Consider these expressions at $t=q^{-1}$, i.e. choose $m=1$. Then,
\be
\begin{array}{cccc}
E_{[4,0]}(t=q^{-1}) &=& \underbrace{x_1^4- \frac{[4]_q }{[3]_q }x_1^3x_2}_{\Psi^{[2,1]}}&-
\underbrace{\frac{q^3 }{[3]_q}x_2^4}_{\Psi^{[1,2]}}\\
E_{[0,5]}(t=q^{-1}) &=&\underbrace{-\frac{1}{q[3]_q} x_1^4x_2}_{\Psi^{[1,2]}}&
\underbrace{ -\frac{ [4]_q  }{q[3]_q  }x_1x_2^4+x_2^5}_{\Psi^{[2,1]}}\\
E_{[5,0]}(t=q^{-1}) &=& \underbrace{x_1^5- \frac{[5]_q }{[4]_q }x_1^4x_2}_{\Psi^{[2,1]}}&-
\underbrace{\frac{q^4 }{[4]_q}x_2^5}_{\Psi^{[1,2]}}\\
E_{[0,6]}(t=q^{-1}) &=& \underbrace{-\frac{1}{q[4]_q} x_1^5x_2
}_{\Psi^{[1,2]}}&
\underbrace{ -\frac{ [5]_q  }{q[4]_q  }x_1x_2^5+x_2^6}_{\Psi^{[2,1]}}\\
\\
&&\ldots&\\
\\
E_{[0,n]}(t=q^{-1}) &=&\underbrace{-\frac{1}{q[n-2]_q} x_1^{n-1}x_2}_{\Psi^{[1,2]}}&
\underbrace{ -\frac{ [n-1]_q  }{q[n-2]_q  }x_1x_2^{n-1}+x_2^n}_{\Psi^{[2,1]}}\\
E_{[n,0]}(t=q^{-1}) &=& \underbrace{x_1^n- \frac{[n]_q }{[n-1]_q }x_1^{n-1}x_2}_{\Psi^{[2,1]}}&-
\underbrace{\frac{q^{n-1} }{[n-1]_q}x_2^n}_{\Psi^{[1,2]}}
\end{array}\nn
\ee
Similarly, putting $t=q^{-2}$ gives rise to
{\footnotesize
\be
\begin{array}{cccc}
E_{[5,0]}(t=q^{-2}) &=& \underbrace{x_1^5- \frac{[2]_q[5]_q }{q[3]_q }x_1^4x_2+{[4]_q[5]_q\over q[2]_q[3]_q}x_1^3x_2^2}_{\Psi^{[2,1]}}&
\underbrace{+{q[5]_q\over[3]_q}x_1x_2^4-\frac{q^3[2]_q }{[3]_q}x_2^5}_{\Psi^{[1,2]}}\\
E_{[0,6]}(t=q^{-2}) &=&\underbrace{-\frac{[2]_q}{q^2[3]_q} x_1^5x_2+\frac{[5]_q}{q^3[3]_q} x_1^4x_2^2}_{\Psi^{[1,2]}}&
\underbrace{+{[4]_q[5]_q\over q^3[2]_q[3]_q}x_1^2x_2^4 -\frac{ [2]_q[5]_q  }{q^2[3]_q  }x_1x_2^5+x_2^6}_{\Psi^{[2,1]}}\\
\\
&&\ldots&\\
\\
E_{[n,0]}(t=q^{-2}) &=& \underbrace{x_1^n- \frac{[2]_q[n]_q }{q[n-2]_q }x_1^{n-1}x_2+{[n-1]_q[n]_q\over q[n-3]_q[n-2]_q}x_1^{n-2}x_2^2}_{\Psi^{[2,1]}}&
\underbrace{+{q^{n-4}[2]_q[n]_q\over[n-3]_q[n-2]_q}x_1x_2^{n-1}-\frac{q^{n-2}[2]_q }{[n-2]_q}x_2^n}_{\Psi^{[1,2]}}\\
E_{[0,n]}(t=q^{-2}) &=&\underbrace{-\frac{[2]_q}{q^2[n-3]_q} x_1^{n-1}x_2+\frac{[2]_q[n-1]_q[n-4]_q}{q^3[n-3]_q} x_1^{n-2}x_2^2}_{\Psi^{[1,2]}}&
\underbrace{+{[n-2]_q[n-1]_q\over q^3[n-4]_q[n-3]_q}x_1^2x_2^{n-2} -\frac{ [2]_q[n-1]_q  }{q^2[n-3]_q  }x_1x_2^{n-1}+x_2^n}_{\Psi^{[2,1]}}
\end{array}\nn
\ee}
i.e. there are just $2!\times (2-1)!=2$ instead of naive $2!\times 2!=4$ independent building blocks.

Looking at these expressions, one can conclude that the non-symmetric Macdonald polynomials are sums of two BA functions,
$E=\Psi^{[2,1]}+\Psi^{[1,2]}$, and $\Psi^{[2,1]}$ is a polynomial of degree $m$, while $\Psi^{[1,2]}$ is a polynomial of degree $m-1$. Moreover, one notes that the formulas look like
\be
E_{[\mu,0]}=\Psi^{[2,1]}(x_1,x_2;\mu)+{1\over x_1}\Psi^{[1,2]}(qx_2,x_1;\mu+1)\nn\\
E_{[0,\mu]}=\Psi^{[1,2]}(x_1,x_2;\mu)+q^{1-\mu}x_2\Psi^{[2,1]}(qx_2,x_1;\mu-1)\nn
\ee
We discuss them in detail in sec.4 (see (\ref{Epsi1}), (\ref{Epsi2})).

\subsection{$N=3$ case}

Now let us see how a similar scheme works in the case of $N=3$. In this case, the non-symmetric Macdonald polynomial $E_\mu$ is still decays into six different pieces associated with separate BA functions at $t=q^{-m}$ so that\footnote{For generic $N$, it has to be $|\mu_i-\mu_{i+1}|>Nm$.} $|\mu_i-\mu_{i+1}|>3m$ for all $i$: otherwise, not all BA functions are well separated. Consider, for instance, the $m=1$ case. Then, one obtains for any $n>3$
\be\label{N3}
E_{[n,0,0]}(t=q^{-1}) &=& \underbrace{x_1^n  -\frac{[n]_q}{[n-1]_q}x_1^{n-1}(x_2+x_3)
 + \frac{[n]_q}{[n-2]_q}x_1^{n-2}x_2x_3}_{\Psi^{([n,0,0],[1,2,3]}+\Psi^{([n,0,0],[1,3,2]}} -
\nn \\
&&\underbrace{- \frac{q^{n-1}}{[n]_q}x_2^n+\frac{q^{n-2}[n]_q}{[n-1]_q[n-2]_q}x_2^{n-1}x_3}_{\Psi^{([n,0,0],[2,1,3]}+\Psi^{([n,0,0],[2,3,1]}}\ \
\underbrace{- \frac{q^{n-1}}{[n]_q}x_3^n+\frac{q^{n-2}[n]_q}{[n-1]_q[n-2]_q}x_2x_3^{n-1}}_{\Psi^{([n,0,0],[3,2,1]}+\Psi^{([n,0,0],[3,1,2]}}\nn\\
\nn\\
E_{[0,n,0]}(t=q^{-1}) &=& \underbrace{  x_2^n -\frac{[n-1]_q}{q[n-2]_q}x_1x_2^{n-1}-\underline{\left(q+\frac{[n-3]_q}{[n-2]_q^2}\right)}x_2^{n-1}x_3
 + \frac{[n-1]_q^2}{q[n-2]_q^2}x_2^{n-2}x_1x_3    }_{\Psi^{([0,n,0],[2,3,1]}+\Psi^{([0,n,0],[2,1,3]}}  - \nn \\
&& \underbrace{ - \frac{q^{n-2}}{[n-2]_q}x_3^n +\frac{q^{n-3}[n-1]_q}{[n-2]_q^2}x_1x_3^{n-1}
 + \frac{q^{n-3}}{[n-2]_q^2}x_2x_3^{n-1}   }_{\Psi^{([0,n,0],[3,2,1]}+\Psi^{([0,n,0],[3,1,2]}} - \nn \\
&&\underbrace{-\frac{1}{q[n-2]_q}x_1^{n-1}x_2+\frac{q^{n-3}}{[n-2]_q^2}x_1^{n-1}x_3
+ \frac{[n-1]_q}{q[n-2]_q^2}x_1^{n-2}x_2x_3   }_{\Psi^{([0,n,0],[1,3,2]}+\Psi^{([0,n,0],[1,2,3]}}
\nn
\ee
\be
E_{[0,0,n]}(t=q^{-1}) &=& \underbrace{x_3^n  -\frac{[n-1]_q}{q[n-2]_q}(x_1+x_2)x_3^{n-1}
 + \frac{[n-1]_q}{q^2[n-3]_q}x_1 x_2x_3^{n-2}}_{\Psi^{([0,0,n],[3,1,2]}+\Psi^{([0,0,n],[3,2,1]}} -\nn
 \\
&&\underbrace{- \frac{1}{q[n-2]_q}x_1^{n-1}x_3+\frac{ [n-1]_q}{q^2[n-2]_q[n-3]_q}x_1^{n-2}x_2x_3}_{\Psi^{([0,0,n],[1,3,2]}+\Psi^{([0,0,n],[1,2,3]}}-\nn\\
&&\underbrace{- \frac{1}{q[n-2]_q}x_2^{n-1}x_3+\frac{ [n-1]_q}{q^2[n-2]_q[n-3]_q}x_2^{n-2}x_1x_3}_{\Psi^{([0,0,n],[2,1,3]}+\Psi^{([0,0,n],[2,3,1]}}\nn
\ee
For large $n$, there is a clear separation into three different pieces, however, we can not distinguish between pairs of $\Psi$'s.

Here we also observe a new phenomenon as compared with the $N=2$ case: the first example of {\bf a non-factorized term} (see an explanation of this in sec.\ref{BR}), which is underlined in this formula.

A full separation into $3!=6$ fragments can be seen at $t=q^{-1}$ for $n>3$ in the next two examples (the requirement $|\mu_i-\mu_{i+1}|>3$ is satisfied at large enough $n$):
{\footnotesize
\be 
E_{[0,n,2n]}(t=q^{-1})=\hspace{14cm}\nn
\ee
\be
\begin{array}{rcll}
 &=&  x_2^nx_3^{2n}\left({\bf 1}
- \frac{[n-1]_q}{q[n-2]_q} \frac{x_1}{x_2}
- \frac{[n - 1]_q}{q[n - 2]_q} \frac{x_2}{x_3}
+ \frac{[2n - 2]_q + q^{n - 2}[n]_q}{q^2[n - 2]_q[2n - 3]_q} \frac{x_1}{x_3}
+\right.&\\
\\
&+&\left.
\frac{[n - 1]_q[2n - 2]_q}{q^2[n - 2]_q[2n - 3]_q}    \frac{ x_1^2}{x_2x_3}
+ \frac{[n - 1]_q[2n - 2]_q}{q^2[n - 2]_q[2n - 3]_q}   \frac{x_1x_2}{x_3^2}
- \frac{[n - 1]_q^2[2n - 2]_q}{q^3[n - 2]_q^2[2n - 3]_q}   \frac{ x_1^2}{x_3^2} \right)- &\Psi^{([0,n,2n],[3,2,1]}
\\ \\
&+& x_2^{2n}x_3^n\left(
- \frac{1}{q[n - 2]_q}\frac{x_3}{x_2}
+ \frac{q^{n - 2} + [2n - 2]_q}{q^2[n - 2]_q[2n - 3]_q}       \frac{ x_1}{x_2}
- \frac{[n - 1]_q[2n - 2]_q}{q^3[n - 2]_q^2[2n - 3]_q} \frac{x_1^2}{x_2^2}
  + \frac{[2n - 2]_q}{q^2[2n - 3]_q[n - 2]_q}    \frac{x_1x_3}{x_2^2}\right)+ &\Psi^{([0,n,2n],[2,3,1]}
  \\ \\
&+&x_1^{2n}x_2^n\left(
- \frac{1}{q^2[2n - 3]_q} \frac{x_3}{x_1}
+ \frac{[n - 1]_q}{q^2[n - 2]_q[2n - 3]_q}  \frac{x_2x_3}{x_1^2 }
+ \frac{[n - 1]_q}{q^2[n - 2]_q[2n - 3]_q}       \frac{x_3^2}{x_1x_2}
- \frac{q^2[n-2]_q^2 + [2n - 2]_q}{q^3[n - 2]_q^2[2n - 3]_q}\frac{ x_3^2}{x_1^2}
\right)+&\Psi^{([0,n,2n],[1,2,3]}
 \\ \\
&+& x_1^{2n}x_3^n\left(
\frac{q^{n - 4}}{[n - 2]_q[2n - 3]_q}\frac{x_2}{x_1}
- \frac{[2n - 2]_q}{q^3[n - 2]_q^2[2n - 3]_q}\frac{x_2^2}{x_1^2}
+ \frac{1}{q^2[n - 2]_q[2n - 3]_q} \frac{x_2x_3}{x_1^2}
\right)+ &\Psi^{([0,n,2n],[1,3,2]}
 \\  \\
&+& x_1^nx_2^{2n}\left(
 \frac{q^{n - 4} }{[n-2]_q [2n - 3]_q}\frac{x_3}{x_2}
+  \frac{ 1}{q^2[n - 2]_q[2n - 3]_q}\frac{x_3^2}{x_1x_2}
- \frac{[2n - 2]_q}{q^3[n - 2]_q^2[2n - 3]_q}  \frac{x_3^2}{x_2^2}
\right)+ &\Psi^{([0,n,2n],[2,1,3]}
 \\   \\
&+&x_1^nx_3^{2n}\left(
- \frac{1}{q[n - 2]_q}\frac{x_2}{x_1}
+ \frac{q^{n - 2} + [2n - 2]_q}{q^2[n - 2]_q[2n - 3]_q} \frac{x_2}{x_3}
+ \frac{[2n - 2]_q}{q^2[n - 2]_q[2n - 3]_q}        \frac{  x_2^2}{x_1x_3}
- \frac{[n - 1]_q[2n - 2]_q}{q^3[n - 2]_q^2[2n - 3]_q}     \frac{  x_2^2}{x_3^2}\right) \ \ \
&\Psi^{([0,n,2n],[3,1,2]}
\end{array}\nn
\ee
}
and
{\footnotesize
\be
E_{0,2n,n}(t=q^{-1})=\hspace{14cm}\nn
\ee
\be
\begin{array}{rcll}
&=& x_2^{2n}x_3^n\left({\bf 1}
- \frac{[n-1]_q}{q[n-2]_q}\frac{x_1}{x_3} - \frac{[n]_q}{[n-1]_q}\frac{x_3}{x_2}
+ \frac{[2n-3]_q^2-[n-2]_q[n-3]_q}{[2n-3]_q[n-1]_q[n-2]_q}\,\frac{x_1}{x_2} -
\right.&
\\
\\
&-&\left.
 \frac{[2n-2]_q[n]_q}{q^2[2n-3]_q[n-2]_q}\frac{x_1^2}{x_2^2}
+ \frac{[2n-2]_q[n-1]_q}{q^2[2n-3]_q[n-2]_q}\frac{x_1^2}{x_2x_3}
+ \frac{[2n-2]_q[n]_q}{q[2n-3]_q[n-1]_q}\frac{x_1x_3}{x_2^2}\right)
- &\Psi^{([0,n,2n],[2,3,1]}  \\  \\
&-&\frac{q^{n-1}}{[n-1]_q} x_2^nx_3^{2n}\left(1
-\frac{[n-1]_q}{q[n-2]_q}\frac{x_1}{x_2}
- \frac{q[2n-3]_q[n-2]_q+[n-3]_q }{q[2n-3]_q[n-2]_q}\frac{x_1}{x_3}
+ \frac{[2n-2]_q[n-1]_q}{q^2[2n-3]_q[n-2]_q}\frac{x_1^2}{x_2x_3}\right)+
&\Psi^{([0,n,2n],[3,2,1]}  \\  \\
&+&x_1^{2n}x_2^n\left(\frac{1}{q^2[n-1]_q[n-2]_q}\frac{x_3}{x_1}
 -\frac{[n]_q}{q^2[2n-3]_q[n-1]_q[n-2]_q}\frac{x_3^2}{x_1^2}
 - \frac{q^{n-3}}{[2n-3]_q[n-2]_q}\frac{x_3^2}{x_1x_2}\right)
 + &\Psi^{([0,n,2n],[1,2,3]} \\  \\
&+&x_1^{2n}x_3^n\left(
-\frac{q^{n-2}[n-1]_q+[n-2]_q^2}{q[2n-3]_q[n-1]_q[n-2]_q}\frac{x_2}{x_1}
-\frac{[n]_q}{q^2[2n-3]_q[n-2]_q}\frac{x_2^2}{x_1^2}
+ \frac{[n-1]_q}{q^2[2n-3]_q[n-2]_q}\frac{x_2^2}{x_1x_3}
+ \frac{[n]_q}{q[2n-3]_q[n-1]_q}\frac{x_2x_3}{x_1^2}
\right)
+ &\Psi^{([0,n,2n],[1,3,2]}
 \\  \\
&+&x_1^n x_2^{2n}\left(
-\frac{1}{q[n-2]_q}\frac{x_3}{x_1} + \frac{[n]_q}{q^2[n-1]_q[n-2]_q}\frac{x_3}{x_2}
- \frac{[2n-2]_q[n]_q}{q^2[2n-3]_q[n-1]_q[n-2]_q}\frac{x_3^2}{x_2^2}
+ \frac{q^{n-1}[n-1]_q^2+[n]_q[n-2]_q}{q[2n-3]_q[n-1]_q[n-2]_q}\frac{x_3^2}{x_1x_2}
\right) +  &\Psi^{([0,n,2n],[2,1,3]}
 \\  \\
&+& x_1^nx_3^{2n}\left(
\frac{q^{n-2}}{[n-1]_q[n-2]_q}\frac{x_2}{x_1}
-\frac{q^{2n-5}[n]_q}{[2n-3]_q[n-1]_q[n-2]_q}\frac{x_2}{x_3}
- \frac{q^{n-3}}{[2n-3]_q[n-2]_q}\frac{x_2^2}{x_1x_3}\right)
&\Psi^{([0,n,2n],[3,1,2]}
\end{array}\nn
\ee
}
One can see that each non-symmetric Macdonald polynomial is a sum of six different pieces, the BA functions, and, in variance with the $N=2$ case, these BA functions are distinct for different polynomials:
\be
E_\mu=\Psi^{(\mu,[1,2,3])}+\Psi^{(\mu,[1,3,2])}+\Psi^{(\mu,[2,1,3])}+\Psi^{(\mu,[2,3,1])}+\Psi^{(\mu,[3,1,2])}+\Psi^{(\mu,[3,2,1])}\nn
\ee
Since there are six different possible orders of $\mu_i$'s in $E_\mu$, one could naively expect $6\times 6=36$ different BA functions. However, this is not the case! Because of the relation
\be
E_{[\mu_2,\ldots,\mu_N,\mu_1+1]}(x_1,x_2,\ldots,x_N)=
q^{-\mu_1}x_NE_{[\mu_1\mu_2,\ldots,\mu_N]}(qx_{n},x_1,x_2,\ldots,x_{N-1})\nn
\ee
the BA functions making up the polynomials $E_{[0,n,2n]}$, $E_{[2n-1,0,n]}$, $E_{[n-1,2n-1,0]}$ are the same (up to permutations and rescalings), and the same is correct for polynomials $E_{[0,2n,n]}$, $E_{[n-1,0,2n]}$, $E_{[2n-1,n-1,0]}$. Hence, there are only $3!\times 2=12$ different BA functions, as many as there are cyclic orbits in the permutation group $S_3$. Since in $S_2$ there is only one orbit, there are only $2!\times 1=2$ independent BA functions, which we observed in the previous subsection.

\section{DAHA system of type $A$}

After describing examples, we go to the systematic study of the non-symmetric triad.

\subsection{Algebra}

\paragraph{DAHA relations.} The $A_{N-1}$ DAHA is parameterized by 2 parameters $q$ and $t$ and consists of elements $T_i$ ($i=1,\ldots,N-1$), $X_j$, $C_j$ ($j=1,\ldots,N$) subject to the relations \cite{Ch,NSCh,BF,dFK1}:

\bigskip

\begin{itemize}
\item At $i=1,\ldots,N-2$ (Hecke algebra):
\be
(T_i-1)(T_i+t^{-1})&=&0\nn\\
\phantom{.}[T_i,T_j]&=&0,\ \ \ \ \ \ \ |i-j|\ge 2\nn\\
T_iT_{i+1}T_i&=&T_{i+1}T_iT_{i+1}\nn
\ee
\item At $i=1,\ldots,N-1$:
\be
tT_iC_{i+1}T_i&=&C_i\nn\\
\phantom{.}[T_i,C_j]&=&0\ \ \ \ \ \ i\ne j,j-1\nn
\ee
\be
tT_iX_iT_i&=&X_{i+1}\nn\\
\phantom{.}[T_i,X_j]&=&0\ \ \ \ \ \ i\ne j,j+1\nn
\ee
\item At $i,j=1,\ldots,N$:
\be
\phantom{.}[C_i,C_j]=0\nn\\
\phantom{.}[X_i,X_j]=0\nn
\ee
and, introducing,
\be
X:=X_1\ldots X_N,\ \ \ \ \ \ \ \ C:=C_1\ldots C_N\nn
\ee
one also adds
\be
X_1C_2=tC_2T_1^2X_1\nn\\
CX_j=qX_jC\nn\\
XC_j=q^{-1}C_jX\nn
\ee
\end{itemize}

\bigskip

\noindent
Now one can define
\be
\pi:=T_{N-1}^{-1}\ldots T_2^{-1}T_1^{-1}C_1\nn
\ee
Then, at $i=1,\ldots,n-2$:
\be
\pi^{-1}T_i\pi=T_{i+1}\nn
\ee
and
\be\label{Cpi}
C_i=t^{N-i}T_iT_{i+1}\ldots T_{N-1}\pi T_1^{-1}T_2^{-1}\ldots T_{i-1}^{-1}
\ee
We will also need an operator
\be
B:=T_{N-1}\ldots T_2T_1 X_1\nn
\ee
such that
\be
C_iB=B C_{i+1}\nn
\ee

\paragraph{$x$-representation.}
This representation is defined to be
\be
X_i F(x_1,x_2,\ldots,x_N):=x_iF(x_1,x_2,\ldots,x_N)\nn
\ee
and the Hecke generators ($i=1,\ldots,N-1$) are realized in this representation as
\be\label{TxA}
T_i=1+{x_i-t^{-1}x_{i+1}\over x_i-x_{i+1}}(\sigma_{i}(x)-1),\ \ \ \ \ i=1,\ldots,N-1
\ee
where $\sigma_{i}(x)$ permutes $x_i$ and $x_{i+1}$.

The element $\pi$ acts in the $x$-representation as
\be\label{pix}
\pi F(x_1,x_2,\ldots,x_N)=F(qx_N,x_1,\ldots,x_{N-1})
\ee
and one can manifestly construct the Cherednik Hamiltonians $C_i$ in the $x$-representation using (\ref{TxA}), (\ref{pix}) and (\ref{Cpi}).

\subsection{Eigenfunctions}

The common {\bf polynomial} eigenfunctions of the Cherednik Hamiltonians $C_i$'s are enumerated by \wcs $\alpha$ (i.e. compositions admitting zero numbers, $\alpha_i\in\mathbb{Z}_{\ge 0}$) on which the symmetric group $S_N$ acts (i.e. the Weyl group for the root system of type $A$), and are called non-symmetric Macdonald polynomials $E_\alpha=E_{w\alpha^+}$, where $\alpha^+$ denotes the Young diagram, i.e. $\alpha^+_1\ge\alpha^+_2\ge\ldots\ge\alpha^+_n\ge 0$, and $w\in S_n$ is the shortest element that produces $\alpha$ from $\alpha^+$:
\be
C_i\cdot E_{\alpha}=\Lambda^{(i)}_{\alpha}\cdot E_{\alpha}\nn
\ee
The eigenvalues are
\be\label{evw}
\Lambda^{(i)}_{\alpha}=q^{\alpha_i}t^{w(\rho)_i}
\ee
where $\rho$ is the shifted Weyl vector with components $N-i$.

The non-symmetric Macdonald polynomials are graded with the grading $|\alpha|:=\sum_{i=1}^n\alpha_i$. They have triangular expansions (hereafter we normalize the polynomials to have unit coefficient of the leading term):
\be\label{nsM}
E_{\alpha}=x^\alpha+\sum_{\beta<\alpha}C_{\alpha\beta}x^\beta
\ee
with the following ordering: for $\alpha=w(\alpha^+)$, $\beta=w'(\beta^+)$, $w,w'\in S_n$, one defines
$\alpha>\beta$ if $\alpha^+>\beta^+$ (e.g., in accordance with the lexicographic order), or, when $\alpha^+=\beta^+$, if the minimal length of $w$ is less than that of $w'$ (Bruhat order \cite{HHL}).

Our goal in this paper is to construct (non-polynomial) eigenfunctions of the Cherednik Hamiltonians at arbitrary eigenvalues.

\section{$N=2$ triad and BA function}

We start with considering $N=2$ case, and construct eigenfunctions of the Cherednik Hamiltonians $C_i$ with arbitrary eigenvalues.

\subsection{$N=2$ triad}

First, we remind that the standard $N=2$ non-symmetric Macdonald polynomials are described by formulas \cite{Mac2}
\be\label{E12}
E_{[\mu_1,\mu_2]}(x_1,x_2)=x_1^{\mu_2}x_2^{\mu_1}\sum_{k=0}\left({x_1\over x_2}\right)^k q^{\mu-k}\ \prod_{j=1}^k{(1-q^{\mu-j+1})\over (1-q^j)}\prod_{j=0}^k{(1-tq^{j})\over(1- tq^{\mu-j})}\nn\\
\ee
when $\mu:=\mu_1-\mu_2\ge 0$, and
\be\label{E21}
E_{[\mu_1,\mu_2]}=x_1^{\mu_1}x_2^{\mu_2}\sum_{k=0}\left({x_1\over x_2}\right)^k
\ \prod_{j=1}^k{(1-q^{\mu-j})\over (1-q^j)}{(1-tq^{j-1})\over (1-tq^{\mu-j})}
\ee
when $\mu:=\mu_2-\mu_1\ge 0$.

The first sums runs up to $\mu$, and the second one, up to $\mu-1$. These expressions admit natural extension to arbitrary complex values of $\mu_{1,2}$ so that the sums become power series instead of polynomials and run up to infinity. However, still it will be two {\bf different} sums, or two branches of one function\footnote{Whether or not these two power series are indeed expansions of a single function on two sheets of a ramified complex plane, is currently an open question.}. Let us mark them by superscripts [2,1] and [1,2] accordingly. These superscripts can be associated with two elements of the permutation group $S_2$ that acts on the set [2,1]: $w={\rm Id}$ and $w=\sigma_1$. It will be also convenient for us to introduce new complex quantities $y_{1,2}=\lambda_{1,2}$ for the extended formulas: $q^{\lambda_i}=q^{\mu_i}t^{\rho_i}$ for the first formula, and $q^{\lambda_i}=q^{\mu_i}t^{-\rho_i}$ for the second one. Here $\rho$ is the Weyl vector with components $\rho_1={1\over 2}$, $\rho_2=-{1\over 2}$. These new variables can be also written by the single formula for the both cases: $y_i=q^{\lambda_i}=q^{\mu_i}t^{w(\rho)_i}$.

Thus, we ultimately will deal with power series
\be\label{NSE}
\mathfrak{E}^{[2,1]}(x_1,x_2;y_1,y_2)&=&{(1-t)\over t(1-q^{\lambda_1-\lambda_2})}
x_1^{\lambda_2}x_2^{\lambda_1}\cdot t^{z_1-z_2\over 2}\cdot\sum_{k=0}\left({x_1\over x_2}\right)^k q^{\lambda_1-\lambda_2-k}\
 \prod_{j=1}^k{(1-tq^{j})\over (1-q^j)}{(1-t^{-1}q^{\lambda_1-\lambda_2-j+1})\over (1-q^{\lambda_1-\lambda_2-j})}\nn\\
\mathfrak{E}^{[1,2]}(x_1,x_2;y_1,y_2)&=&x_1^{\lambda_1}x_2^{\lambda_2}\cdot t^{z_1-z_2\over 2}\cdot\sum_{k=0}\left({x_1\over x_2}\right)^k \
\prod_{j=1}^k{(1-tq^{j-1})\over (1-q^j)}{(1-t^{-1}q^{\lambda_2-\lambda_1-j})\over (1-q^{\lambda_2-\lambda_1-j})}
\ee

These power series admit {\bf two} different polynomial reductions. The first reduction $y_i=q^{\lambda_i}=q^{\mu_i}t^{w(\rho)_i}$ with proper $\mu_i$'s returns them to the non-symmetric Macdonald polynomials.

There is also another reduction $t=q^{-m}$, $m\in\mathbb{Z}_{\ge 0}$, which produces the Baker-Akhiezer functions, and cuts the first sum to $m$ terms, and the second one, to $m+1$ terms producing (quasi)polynomials:
\be\label{BAE}
\Psi^{[2,1]}_m(x_1,x_2;\lambda_1,\lambda_2)&=&{(1-t)\over t(1-q^\lambda)}
x_1^{\lambda_2}x_2^{\lambda_1}\cdot t^{z_1-z_2\over 2}\cdot\sum_{k=0}^{m-1}\left({x_1\over x_2}\right)^k q^{\lambda_1-\lambda_2-k}\ \prod_{j=1}^k{(1-q^{j-m})\over (1-q^j)}{(1-q^{\lambda_1-\lambda_2-j+m+1})\over (1-q^{\lambda_1-\lambda_2-j})}\nn\\
\Psi^{[1,2]}_m(x_1,x_2;\lambda_1,\lambda_2)&=&x_1^{\lambda_1}x_2^{\lambda_2}\cdot t^{z_1-z_2\over 2}\cdot\sum_{k=0}^m\left({x_1\over x_2}\right)^k \
\prod_{j=1}^k{(1-q^{j-m-1})\over (1-q^j)}{(1-q^{\lambda_2-\lambda_1-j+m})\over (1-q^{\lambda_2-\lambda_1-j})}
\ee

One can generate the non-symmetric Macdonald polynomials at this value\footnote{Note that this reduction can make the Macdonald polynomial singular. It does not happen if $m>\mu_1-\mu_2$ or $m\le{\mu_1-\mu_2\over 2}$ for $E^{[2,1]}_{[\mu_1,\mu_2]}(x_1,x_2)$, and $[\ldots]$ denotes the integer part. Similarly, it does not happen if $m\ge\mu_1-\mu_2$ or $m\le{\mu_1-\mu_2-1\over 2}$ for $E^{[1,2]}_{[\mu_2,\mu_1]}(x_1,x_2)$. In order to deal with the any $m$, one has to change the normalization of the Macdonald polynomial: one has to normalize them correspondingly with the factors (\ref{norm}) below.} of parameter $t$ from the BA functions:
\be\label{Epsi1}
E^{[2,1]}_{[\mu_1,\mu_2]}(x_1,x_2)=\Psi^{[2,1]}_m\Big(x_1,x_2;\mu_1-{m\over 2},\mu_2+{m\over 2}\Big)+{q^{-\mu_2}\over x_1}\Psi^{[1,2]}_m\Big(qx_2,x_1;\mu_2+{m\over 2},\mu_1+1-{m\over 2}\Big)
\ee
and we put here $m>\mu_1-\mu_2$.

Similarly,
\be\label{Epsi2}
E^{[1,2]}_{[\mu_1,\mu_2]}(x_1,x_2)=\Psi^{[1,2]}_m\Big(x_1,x_2;\mu_1+{m\over 2},\mu_2-{m\over 2}\Big)+q^{1-\mu_2}x_2\Psi^{[2,1]}_m\Big(qx_2,x_1;\mu_2-1-{m\over 2},\mu_1+{m\over 2}\Big)
\ee
We already observed these formulas in sec.2.1.

From these two relations immediately follows
\be\label{Cyc}
E^{[1,2]}_{\mu_1,\mu_2+1}(qx_2,x_1) = q^{\mu_1} x_1 E^{[2,1]}_{\mu_2,\mu_1}(x_1,x_2)
\ee
since
\be
\Psi_m(q\vec x;\vec\lambda)=q^{\sum_i\lambda_i}\Psi_m(\vec x;\vec\lambda)\nn
\ee
Thus, one can obtain the non-symmetric polynomial either directly from the universal power series (\ref{NSE}) or as a sum of two BA functions at $t=q^{-m}$. Hence, just as in the symmetric case, the two reductions are not permutable (see a similar example in the symmetric case in \cite{MMP3}).

\subsection{Properties of the BA functions}

\paragraph{Symmetricity.} Consider differently normalized Baker-Akhiezer functions:
\be\label{norm}
\Psi^{[2,1]}_m(x_1,x_2;\lambda_1,\lambda_2)\to \prod_{j=0}^{m-1}\Big(q^{{\lambda_1-\lambda_2\over 2}-j}-q^{\lambda_2-\lambda_1\over 2}\Big)\Psi^{[2,1]}_m(x_1,x_2;\lambda_1,\lambda_2)\nn
\\
\Psi^{[1,2]}_m(x_1,x_2;\lambda_1,\lambda_2)\to \prod_{j=1}^{m}\Big(q^{{\lambda_1-\lambda_2\over 2}-j}-q^{\lambda_2-\lambda_1\over 2}\Big)\Psi^{[1,2]}_m(x_1,x_2;\lambda_1,\lambda_2)
\ee
Similarly to the symmetric case, these BA functions are symmetric (equivariant) with respect to the permutation\footnote{In the symmetric case changing $x_i$ for inverse is not necessary because of the symmetries of the BA functions.} of $x_{1,2}^{-1}$ and $y_{1,2}$:
\be
\Psi_m(x_1^{-1},x_2^{-1};y_1,y_2)=\Psi_m(y_1^{-1},y_2^{-1};x_1,x_2)\nn
\ee

\paragraph{Defining relations for BA functions.} Formula (\ref{Cyc}) is one of the relations of the Knop-Sahi recursion \cite{KS,HHL}. The other one is
\be
E^{[2,1]}_{[\mu_1,\mu_2]}(x_1,x_2)=
{(tx_1-x_2)\over t(x_1-x_2)} E^{[1,2]}_{[\mu_2,\mu_1]}(x_2,x_1)+{1\over t}
{(t-1)\over (tq^{\mu_1-\mu_2}-1)}{(tq^{\mu_1-\mu_2}x_2-x_1)\over(x_2-x_1)} E^{[1,2]}_{[\mu_2,\mu_1]}(x_1,x_2)\nn
\ee
This relation is associated with the defining conditions for the BA functions, which is a counterpart of the periodicity condition in the case of symmetric triad:
\be\label{BAc}
\Psi^{[2,1]}_m(x,1;\lambda_1,\lambda_2)-{(q^m-x)\over(1-x)}q^{1-\lambda_1-{m\over 2}}x\Psi^{[2,1]}_m(qx,1;\lambda_1-1,\lambda_2)
-{(q^m-1)(x-q^{\lambda_1-\lambda_2})\over(1-q^{\lambda_1-\lambda_2})(1-x)}
\Psi^{[1,2]}_m(x,1;\lambda_2,\lambda_1)=0
\ee
\be
{q^{\lambda_1+{m\over 2}}\over x}\Psi^{[1,2]}_m(1,x;\lambda_2,\lambda_1+1)
-{(q^m-qx)\over(1-qx)}\Psi^{[1,2]}_m(1,qx;\lambda_2,\lambda_1)-\nn\\
-{(q^m-1)(qx-q^{\lambda_1-\lambda_2})\over(1-q^{\lambda_1-\lambda_2})(1-qx)}
q^{-\lambda_2-{m\over 2}}\Psi^{[2,1]}_m(1,x;\lambda_1-1,\lambda_2)=0\nn
\ee
If one looks for a solution of (\ref{BAc}) in the (quasi)polynomial form of
\be
\Psi^{[2,1]}_m(x_1,x_2;\lambda_1,\lambda_2)&=&
x_1^{\lambda_2}x_2^{\lambda_1}\cdot q^{m(z_2-z_1)\over 2}\cdot\sum_{k=0}^{m-1}c^{[2,1]}_k(\lambda_1,\lambda_2;q,m)\left({x_1\over x_2}\right)^k\nn\\
\Psi^{[1,2]}_m(x_1,x_2;\lambda_1,\lambda_2)&=&x_1^{\lambda_1}x_2^{\lambda_2}\cdot q^{m(z_2-z_1)\over 2}\cdot\sum_{k=0}^m
c^{[1,2]}_k(\lambda_1,\lambda_2;q,m)\left({x_1\over x_2}\right)^k\nn
\ee
the solution to (\ref{BAc}) is unique up to normalization.

\subsection{Universal solution}

Note that neither $\mathfrak{E}_m\Big(x_1,x_2;\lambda_1,\lambda_2\Big)$, nor $\Psi_m(x_1,x_2;\lambda_1,\lambda_2)$ are eigenfunctions of the Cherednik Hamiltonians (\ref{Ch}).
This is because the universal power series are the series in powers of ${x_1\over x_2}$, and the Cherednik Hamiltonians relate expansions in ${x_1\over x_2}$ and in ${x_2\over x_1}$ since they contain the permutation operator $\sigma_1$. In order to construct eigenfunctions with arbitrary eigenvalues, we use the combinations (\ref{Epsi1}), (\ref{Epsi2}):
\be
{\cal E}^{[2,1]}(x_1,x_2;\lambda_1,\lambda_2)=\mathfrak{E}^{[2,1]}_m\Big(x_1,x_2;\lambda_1,\lambda_2\Big)+{q^{-\lambda_2}\over t^{1\over 2}x_1}\mathfrak{E}^{[1,2]}_m\Big(qx_2,x_1;\lambda_2,\lambda_1+1\Big)\nn\\
{\cal E}^{[1,2]}(x_1,x_2;\lambda_1,\lambda_2)=\mathfrak{E}^{[1,2]}_m\Big(x_1,x_2;\lambda_1,\lambda_2\Big)+t^{-{1\over 2}}q^{1-\lambda_2}x_2\mathfrak{E}^{[2,1]}_m\Big(qx_2,x_1;\lambda_2-1,\lambda_1\Big)\nn
\ee
and similarly for the BA functions.

These combinations {\bf are} the eigenfunctions with the same eigenvalues for ${\cal E}^{[2,1]}$ and ${\cal E}^{[1,2]}$:
\be
C_i\cdot{\cal E}(x_1,x_2;\lambda_1,\lambda_2)=q^{\lambda_i}t^{1\over 2}\ {\cal E}(x_1,x_2;\lambda_1,\lambda_2)\nn
\ee
and, hence, provide us with the universal solution to the Cherednik integrable system.

\subsection{Another realization of $N=2$ case}

The results of the previous section can be looked at in another way: one can use the relation
\be
\prod_{j=1}^{\lambda-k}{(1-q^{\lambda-j+1})\over (1-q^j)}\prod_{j=0}^{\lambda-k}{(1-tq^{j})\over(1- tq^{\lambda-j})}=
\prod_{j=1}^k{(1-q^{\lambda-j+1})\over (1-q^j)}{(1-tq^{j-1})\over (1-tq^{\lambda-j+1})}\nn
\ee
in order to represent the non-symmetric Macdonald polynomials at $N=2$ in the form
\be\label{E12a}
E^{[2,1]}_{[\mu_1,\mu_2]}(x_1,x_2)=x_1^{\mu_1}x_2^{\mu_2}\sum_{k=0}\left({x_2\over x_1}\right)^k q^k\ \prod_{j=1}^k{(1-q^{\mu-j+1})\over (1-q^j)}{(1-tq^{j-1})\over(1- tq^{\mu-j+1})}\nn\\
\ee
where $\mu:=\mu_1-\mu_2\ge 0$, and
\be\label{E21a}
E^{[1,2]}_{[\mu_1,\mu_2]}={1-t\over 1-tq^{\mu-1}}x_1^{\mu_2-1}x_2^{\mu_1+1}\sum_{k=0}\left({x_2\over x_1}\right)^k
\ \prod_{j=1}^k{(1-q^{\mu-j})\over (1-q^j)}{(1-tq^{j})\over (1-tq^{\mu-j-1})}
\ee
where $\mu:=\mu_2-\mu_1\ge 0$. It gives rise to another pair of the NS power series:
{\footnotesize\be\label{NSEa}
{\mathfrak{E}}^{([2,1],[1,2])}(x_1,x_2;y_1,y_2)&=&
x_1^{\lambda_1}x_2^{\lambda_2}\cdot t^{z_2-z_1\over 2}\cdot\sum_{k=0}\left({x_2\over x_1}\right)^k q^k\ \prod_{j=1}^k{(1-t^{-1}q^{\lambda-j+1})\over (1-q^j)}{(1-tq^{j-1})\over(1- q^{\lambda-j+1})},\ \ \ \ \ \ \ \ \lambda:=\lambda_1-\lambda_2\nn\\
{\mathfrak{E}}^{([1,2],[1,2])}(x_1,x_2;y_1,y_2)&=&{1-t\over 1-q^{\lambda-1}}x_1^{\lambda_2-1}x_2^{\lambda_1+1}\cdot t^{z_2-z_1\over 2}\cdot\sum_{k=0}\left({x_2\over x_1}\right)^k
\ \prod_{j=1}^k{(1-t^{-1}q^{\lambda-j})\over (1-q^j)}{(1-tq^{j})\over (1-q^{\lambda-j-1})},\ \ \ \ \ \ \ \lambda:=\lambda_2-\lambda_1\nn\\
\ee}
and to the corresponding other pair of the BA functions at $t=q^{-m}$. We use here the second superscript $[2,1]$ in order to stress that the power series is a series of ${x_2\over x_1}$, i.e. $x_2<x_1$. Hence, the power series of the previous subsections of this section are ${\mathfrak{E}}^{([2,1],[1,2])}$ and ${\mathfrak{E}}^{([1,2],[1,2])}$.

Hence, there are totally 4 different NS functions and 4 different BA functions. Now one constructs eigenfunctions of the Cherednik Hamiltonians with arbitrary eigenvalues $q^{\lambda_i}t^{1\over 2}$ in terms of these NS functions:
\be
{\cal E}^{[2,1]}(x_1,x_2;\lambda_1,\lambda_2)=\mathfrak{E}^{([2,1],[1,2])}\Big(x_1,x_2;\lambda_1,\lambda_2\Big)+
{\mathfrak{E}}^{([2,1],[2,1])}\Big(x_1,x_2;\lambda_1,\lambda_2\Big)\nn\\
{\cal E}^{[1,2]}(x_1,x_2;\lambda_1,\lambda_2)=\mathfrak{E}^{([1,2],[1,2])}\Big(x_1,x_2;\lambda_1,\lambda_2\Big)+
{\mathfrak{E}}^{([1,2],[2,1])}\Big(x_1,x_2;\lambda_1,\lambda_2\Big)\nn
\ee

Note that
\be\label{rels}
{\mathfrak{E}}^{([2,1],[1,2])}(x_1,x_2;y_1,y_2)={q^{-\lambda_2}t^{-{1\over 2}}\over x_1}\mathfrak{E}^{([1,2],[2,1])}(qx_2,x_1;y_2,qy_1)\nn\\
{\mathfrak{E}}^{([1,2],[1,2])}(x_1,x_2;y_1,y_2)={q^{-\lambda_2}t^{1\over 2}\over x_1}\mathfrak{E}^{([2,1],[2,1])}(qx_2,x_1;y_2,qy_1)
\ee
This leaves only two independent functions of four, these are the functions we dealt with in the previous section. The reason for relations (\ref{rels}) is the cyclic symmetry (\ref{Cyc}).

This approach is immediately extended to arbitrary $N$: one has $N!$ different NS functions associated with $N!$ different branches of non-symmetric Macdonald polynomials {\bf and} each of these branches is associated with $N!$ different power series (they are power series in different ratios). Thus, totally, one can enumerate the NS function by {\bf two} elements $w$ and $v$, and cyclic symmetry in this case leaves $N!(N-1)!$ independent functions out of these $(N!)^2$.

\section{Generic non-symmetric triad}

Now we describe how the consideration of the previous section is extended to the case of generic $N$.

\subsection{Branching rules\label{BR}}

In the previous paper \cite{NSM5} of the series devoted to construction of eigenfunctions of the Cherednik Hamiltonians with arbitrary eigenvalues, we considered branching rules for the non-symmetric Macdonald polynomials, which allows us to introduce a notion of non-symmetric triad. We briefly repeat the main steps here.

\begin{itemize}
\item
Our main object was the skew non-symmetric polynomials $E_{\mu/\nu}(\vec y)$  defined via the expansion
\be\label{recE}
E_\mu(x_1,\ldots,x_{k-1},y,x_{k+1},\ldots,x_{N})&=&\sum_{\nu:\ l_\nu\le N} y^{|\mu|-|\nu|}E_{\mu/\nu}^{(k)}E_\nu(x_1,\ldots,x_{k-1},x_{k+1},\ldots,x_N)
\ee
The sum here runs over the weak compositions $\nu$ with $N-1$ elements, giving rise to the recursion $N$.

\item In \cite{NSM5}, we also realized that the skew coefficients $E_{\mu/\nu}^{(k)}$'s are factorized (!), and provided a set of explicit expressions for $E_{\mu/\nu}^{(k)}$.

\item Note that $E_{\mu/\nu}^{(k)}$ at different $k$ are related with each other. To understand this, we use that
the non-symmetric Macdonald polynomials enjoy the property (Knop–Sahi recurrence) \cite{KS,HHL}
\be\label{B}
E_{[\mu_2,\ldots,\mu_N,\mu_1+1]}(x_1,x_2,\ldots,x_N)=
q^{-\mu_1}x_NE_{[\mu_1\mu_2,\ldots,\mu_N]}(qx_{n},x_1,x_2,\ldots,x_{N-1})
\ee
From this formula and (\ref{recE}), it follows that
\be
E_{[\mu_1,\ldots,\mu_N]}(x_1,x_2,\ldots,x_N)&=&q^{\mu_1+1}x_1^{-1}
E_{[\mu_2,\ldots,\mu_N,\mu_1+1]}(x_2,\ldots,x_N,q^{-1}x_1)=\nn\\
&=&q^{\mu_1+1}x_1^{-1}\sum_{\nu:\ l_\nu\le N-1} (q^{-1}x_1)^{\mu|-|\nu|+1}
E_{[\mu_2,\ldots,\mu_N,\mu_1+1]/\nu}^{(N)}E_{\nu}(x_2,\ldots,x_N)=\nn\\
&=&q^{|\nu|-|\mu|+\mu_1}
\sum_{\nu:\ l_\nu\le N-1}x_1^{|\mu|-|\nu|} E_{[\mu_2,\ldots,\mu_N,\mu_1+1]/\nu}^{(N)}E_{\nu}(x_2,\ldots,x_N)\nn
\ee
i.e.
\be
E^{(1)}_{\mu/\nu}=q^{|\nu|-|\mu|+\mu_1}E_{[\mu_2,\ldots,\mu_N,\mu_1+1]/\nu}^{(N)}\nn
\ee
One can similarly obtain all other $E^{(k)}_{\mu/\nu}$ using other moves from the Knop-Sahi recurrence \cite{KS,HHL}.
\end{itemize}

The origin of formulas for the BA functions from the skew non-symmetric polynomial expansion explains the phenomenon that emerge at higher $N>2$: the coefficients in front of degrees of ratios of $x_i$'s are sometimes not factorized. We saw an example of this in formula (\ref{N3}) (the underlined term), and, for instance, in the case of $N=3$, an explicit formula for one of the BA functions (that associated with the power series in ${x_1\over x_2}$, ${x_1\over x_3}$, ${x_2\over x_3}$, i.e. $x_3>x_2>x_1$) is
\be
\Psi_1^{([3,2,1],[3,2,1])}(x;\mu+\rho)=x_1^{\mu_3}x_2^{\mu_2}x_3^{\mu_1}\left(\psi_{00}+\psi_{01}{x_2\over x_3}+\psi_{10}{x_1\over x_2}+\psi_{11}{x_1\over x_3}\right)\nn
\ee
where
\be
\psi_{00}&=&-q^{\mu_1+\mu_2-2\mu_3-3}
{1\over[\mu_1-\mu_2-1]_q[\mu_2-\mu_3-1]_q^2}-q^{\mu_1-\mu_3-1}{[\mu_2-\mu_3-2]_q[\mu_2-\mu_3]_q\over [\mu_1-\mu_3-2]_q[\mu_2-\mu_3-1]_q^2}\nn\\
\psi_{01}&=&q^{\mu_1-\mu_3-2}{[\mu_1-\mu_2]_q\over[\mu_1-\mu_2-1]_q[\mu_1-\mu_3-2]_q}\nn\\
\psi_{10}&=&q^{\mu_1-\mu_3-2}{[\mu_2-\mu_3]_q\over [\mu_1-\mu_3-2]_q[\mu_2-\mu_3-1]_q}\nn\\
\psi_{11}&=&-q^{\mu_1-\mu_3-3}{[\mu_1-\mu_2]_q[\mu_2-\mu_3]_q\over[\mu_1-\mu_2-1]_q[\mu_1-\mu_3-2]_q[\mu_2-\mu_3-1]_q}\nn
\ee
The first term in $\psi_{00}$ comes from $E_{\mu/\nu}$ where $\nu$ is a Young diagram, all other terms come from $E_{\mu/\nu}$ where $\nu$ is {\bf not} a Young diagram. This is the reason for non-factorizability.

\subsection{Extension to non-symmetric triad}

\subsubsection{Different Weyl chambers}

Let us note that one can repeat the recursion in $N$: $N\to N-1\to\ldots\to 2\to 1$, and obtain for the non-symmetric Macdonald polynomial an explicit formula of the form
\be\label{Er}
E_\mu(x_1,x_2,\ldots,x_N)=x_1^{|\mu|}\sum_{\{\nu_k\}}\prod_{k=1}^{N-1}\left({x_{k+1}\over x_k}\right)^{|\nu_k|}E^{(k)}_{\nu_{k}/\nu_{k-1}}
\ee
Here we put $\nu_0:=\mu$, the sum runs over weak compositions $\nu_i$'s, and we have chosen at each step of the recursion expanding w.r.t. to $y=x_i$ with the smallest number, i.e. successively w.r.t. to $x_1$, $x_2$, etc.

The coefficients in formula (\ref{Er}) are ratios of $q$-Pochhammer symbols \cite{NSM5}, and (\ref{Er}) admits an immediate extension to arbitrary complex values of elements of $\mu_i$, $i=1,\ldots,N$: $E_\mu\to \mathfrak{E}(\vec\mu)$. However, the sum in this case runs up to infinity, and the polynomial becomes a power series of ratios ${x_{k-1}\over x_k}$, $k=2,\ldots,N$. This concrete set of ratios is due to our choice of the recursion, and is associated with a concrete Weyl chamber. We denote it $s_0=[N,\ldots,1]$ in accordance with the order of $x_i$'s: $x_N> x_{N-1}>\ldots > x_1$.
One could choose any other Weyl chamber, and any other set of ratios. Each set $s$, and each Weyl chamber can be labeled by a permutation $v$ from the symmetric (Weyl) group $S_N$, $s=v(s_0)$. Hence, one has to label the power series $\mathfrak{E}^v(\vec\mu)$ by the permutation $v$.

In variance with the case of weak composition $\mu$, when all $\mathfrak{E}^v(\vec\mu)$ with different $v$'s are equal to each other, the power series are absolutely distinct: they are different power series of distinct ratios.

\subsubsection{Different branches}

The second essential point concerning the extension of formula (\ref{Er}) to arbitrary complex $\mu$ is that, in the case of weak composition $\mu$, the formula essentially depends on the order of entries in the weak composition: each permutation producing a new composition gives rise to another branch of the non-symmetric Macdonald polynomial, see examples in \cite{NSM5}. For instance, in the case of $N=3$, there are six possible orders of numbers $\mu_{1,2,3}$, and the order $\mu_1\ge\mu_2\ge\mu_3$ is associated with the sequence $[3,2,1]$, while all other are obtained by six permutations of $[3,2,1]$. Note that, if, say, $\mu_1<\mu_2=\mu_3$, we write $[1,3,2]$: when two numbers are equal to each other, we assume the left one corresponds to a larger number in the sequence of first $N$ integers. A priori it could have happened that boundary cases not fit together with either $\mu_1<\mu_2<\mu_3$ chamber or $\mu_1<\mu_3<\mu_2$ chamber, instead forming domain walls. Wether this is a general phenomenon of a peculiarity of the $A_N$ root system remains to be studied.

Thus, there are $N!$ different branches of $E$ also enumerated by compositions $w$. We will choose as the reference branch, i.e. that with $w=\Id$, the one associated with $\mu$ that is a Young diagram.

Hence, one ultimately comes to the power series which is labeled by two permutations: $\mathfrak{E}^{(w,v)}(\vec x;\vec y)$. This is what we call {\bf non-symmetric triad}, which has a set of properties much similar the case of symmetric triad.
Each triad is a power series of $N$ variables $x_i$ and of $N$ parameters $y_i$. We will also use the notation $x_i:=q^{z_i}$, $y_i:=q^{\lambda_i}$. Now we describe the basic properties of the non-symmetric triad.

\subsection{Properties of non-symmetric triad}

The triad has two polynomial reductions.

\begin{itemize}
\item
{\bf Polynomial reduction: non-symmetric Macdonald polynomial.}
The first polynomial reduction is achieved by imposing on $\mathfrak{E}^{(w,v)}(x_1,\ldots,x_N;y_1,\ldots,y_N;q,t)$ the condition $y_i=q^{\mu_i}t^{w(\rho)_i}$, where $\mu$ is a set of non-negative integers ordered in accordance with \footnote{For instance, $w=\Id$ means that $\mu_1\ge\mu_2\ge\mu_3$.} $w(s_0)$, $w$ acts on the sequence $(\rho_1,\ldots,\rho_N)$, and $\rho$ is the Weyl vector: $\rho_i={1\over 2}(N-2i+1)$. This gives rise to the non-symmetric Macdonald polynomial $E_\mu(\vec x;q,t)$, and it does not depend on the Weyl chamber $v$, while the dependence on $w$ preserves in the order of integers in the weak composition $\mu$:
\be
\mathfrak{E}^{(w,v)}(\vec x;\{q^{\mu_i}t^{w(\rho)_i}\};q,t)=E_\mu(\vec x;q,t)\nn
\ee

\item
{\bf Polynomial reduction: Baker-Akhiezer function.}
The second polynomial reduction gives rise to the (quasi)polynomial Baker-Akhiezer (BA) function upon a special choice of parameter $t=q^{-m}$, $m\in\mathbb{Z}_{\ge 0}$:
\be
\mathfrak{E}^{(w,v)}(\vec x;\vec y;q,q^{-m})=\Psi^{(w,v)}_m(\vec x;\vec y;q)\nn
\ee
In this case, the infinite power series also becomes a finite sum, a (quasi)polynomial of the degree, at most, $m$. To be exact,
\be\label{BA}
\Psi^{(w,v)}_m(v^{-1}(\vec x);\vec y;q)=x^{w(\lambda)}t^{-z\cdot \rho}
\sum_{\{m\ge k_{ij}\ge 0\}}\prod_{i<j}\left({x_i\over x_j}\right)^{k_{ij}}\psi^{(w,v)}(\{k_{ij}\};\{y_i\};q)
\ee
where $\psi^{(w,v)}(\{k_{ij}\};\{y_i\};q)$ are some coefficients.

This reduction is done for all power series at once, while the first reduction is made for each component separately: the reduction depends on $w$ enumerating the vector components.
On the other hand, this reduction depends both on $w$ and on $v$.

Note that, in contrast with the first reduction, the BA function depends on the Weyl chamber, and one can reproduce the non-symmetric Macdonald polynomial at $t=q^{-m}$ summing the BA functions over the Weyl chambers:
\be
E_\mu(\vec x;q,q^{-m})=\sum_{v\in S_N}\Psi^{(w,v)}_m(\vec x;\{q^{\mu_i}t^{w(\rho)_i}\};q)\nn
\ee
Thus, one can obtain the non-symmetric Macdonald polynomial either from $\mathfrak{E}^{(w,v)}(x_1,\ldots,x_N;y_1,\ldots,y_N;q,t)$ and then choosing $t=q^{-m}$, or as a sum of the BA functions, if the choice $t=q^{-m}$ is made first. This is due to non-permutability of limits, see \cite{MMP3}.

This connects the two polynomial reductions. At large enough differences between $\mu_i$ as compared with $m$, the non-symmetric Macdonald polynomial parts into well-separated $N!$ pieces, each of them being proportional to the corresponding BA function.

\item
{\bf Cherednik eigenfunction.}
The Cherednik Hamiltonians $C_i$ contain the permutation operators $\sigma_i$ that permute $x_i$ and $x_{i+1}$. This means that any single power series $\mathfrak{E}^{(w,v)}(\vec\mu)$ can not be their eigenfunction. We claim that the sum
\be\label{lc}
{\cal E}^{w}(\vec x;\vec y;q,t)=\sum_{v\in W}\mathfrak{E}^{(w,v)}(\vec x;\vec y;q,t)
\ee
becomes an eigenfunction:
\be\label{ee}
C_i\cdot {\cal E}^w(x_1,\ldots,x_N;y_1,\ldots,y_N;q,t)
=t^{N-1\over 2}
y_i\cdot {\cal E}^w(x_1,\ldots,x_N;y_1,\ldots,y_N;q,t)
\ee
Thus, there are totally $N!$ eigenfunctions with arbitrary eigenvalues $\vec y$, which are associated with $N!$ different branches of the non-symmetric triad. Following \cite{dFK2}, we call these eigenfunctions {\bf universal solutions}.

Note that under the first reduction, the obtained non-symmetric Macdonald polynomial {\bf is} a solution to the eigenvalue equations (\ref{ee}) without any additional sum, since all Weyl chambers give rise to the same answer.
\end{itemize}

\subsection{Relations between different triads}

Note that different branches of the triad are related by formulas originated from the Knop-Sahi recursion \cite{KS,HHL}.

\subsubsection{Cyclic permutation}

First of all, there is a direct relation inherited from (\ref{B}):
\be\label{Cyc3}
{\cal E}^{v_cw}(x_1,x_2,\ldots,x_N;y_2,\ldots,y_N,qy_1;q,t)=
t^{-w(\rho)_1}y_1x_N{\cal E}^w(qx_{n},x_1,x_2,\ldots,x_{N-1};y_1,\ldots,y_N;q,t)
\ee
where $v_c$ denotes the cyclic permutation $[1,2,\ldots,N]\to[2,\ldots,N,1]$. It induces a set of relations between different branches of non-symmetric triads like
\be
\mathfrak{E}^{(v_cw,v)}(x_1,x_2,\ldots,x_N;y_2,\ldots,y_N,qy_1;q,t)=
t^{-w(\rho)_1}y_1x_N\mathfrak{E}^{(w,v_c^{-1}v)}(qx_{n},x_1,x_2,\ldots,x_{N-1};y_1,\ldots,y_N;q,t)\nn
\ee
In fact, one can repeat this cyclic permutation $N$ times to express all the branches belonging to the same cyclic orbit through those of one representative. This means that out of $(N!)^2$ power series $\mathfrak{E}^{(w,v)}(\vec x;\vec y)$ only $N!(N-1)!$ becomes distinct power series.

\subsubsection{DAHA generators\label{5.4.2}}

There is a natural action of the Weyl group $S_N$ on different branches of ${\cal E}^w$. It is given by generators $T_i$ from the $A_{N-1}$ DAHA algebra:
\be
\begin{array}{rcll}
T_i{\cal E}^w(\vec x;\vec y;q,t)&=&C^{(1)}_i{\cal E}^w(\vec x;\vec y;q,t)+{\cal E}^w(\vec x;\sigma_i(\vec y);q,t),\ \ \ \ \ \ &\hbox{if}\ \ \ \ \ w(s_0)_i<w(s_0)_{i+1}\cr\cr
T_i{\cal E}^w(\vec x;\vec y;q,t)&=&C^{(1)}_i{\cal E}^w(\vec x;\vec y;q,t)+C^{(2)}_i{\cal E}^w(\vec x;\sigma_i(\vec y);q,t),\ \ \ \ \ \ &\hbox{if}\ \ \ \ \ w(s_0)_i>w(s_0)_{i+1}
\end{array}\nn
\ee
where
\be
C^{(1)}_{i,\alpha}:=-{(1-t)y_i\over t\left(y_i-y_{i+1}\right)}\ \ \ \ \ \ \ \ \ \
C^{(2)}_{i,\alpha}:={\left(y_i-ty_{i+1}\right)\left(y_i-t^{-1}y_{i+1}\right)\over t\left(y_i-y_{i+1}\right)^2}\nn
\ee
Because of (\ref{TxA}), 
it implies the relations
\be\label{KS2}
{(t-1)(y_{i+1}x_{i+1}-y_ix_i)\over(y_{i+1}-y_i)}{\cal E}^w(\vec x;\vec y;q,t)=t(x_{i+1}-x_i){\cal E}^w(\vec x;\sigma_i(\vec y);q,t)
+(tx_i-x_{i+1}){\cal E}^w(\sigma_i(\vec x);\vec y;q,t)
\ee
at $w(s_0)_i>w(s_0)_{i+1}$.

These relations again can be translated to the relations between different branches of non-symmetric triads. At $t=q^{-m}$ they are also translated to the set of linear equations that unambiguously (up to normalization) fix the coefficients $\psi^{(w,v)}(\{k_{ij}\};\{y_i\};q)$ of the Baker-Akhiezer functions. In other words, they substitute the periodicity conditions due to O. Chalykh \cite{Cha}, which fix the BA function in the symmetric case.

\subsection{Properties of the BA function}

At $t=q^{-m}$, $m\in\mathbb{Z}_{\ge 0}$, the non-symmetric triad becomes the (quasi)polynomial Baker-Akhiezer function. It celebrates a set of properties that are quite similar to those in the symmetric case. These are:
\begin{itemize}
\item If one uses the anzatz (\ref{BA}) for the BA function, the set of equations (\ref{Cyc3}), (\ref{KS2}) gives rise to the system of linear equations unambiguously (up to normalization) determining the coefficients $\psi^{(w,v)}(\{k_{ij}\};\{y_i\};q)$ and, therefore, the BA functions.
\item With a properly chosen normalization, the BA function is symmetric with respect to the permutation of $\vec x^{-1}$ and $\vec y$:
\be
\Psi_m^{(w,v)}(\vec x^{-1};\vec y)=\Psi_m^{(w,v)}(\vec y^{-1};\vec x)\nn
\ee
In particular, it results in the notorious duality of the non-symmetric Macdonald polynomials\footnote{Note that in the symmetric case changing $x_i$ for inverse was not necessary because of the symmetries of the BA functions.} \cite{CherednikConj}
\be 
{E_\alpha\left({1\over{\Lambda}^{(i)}_\beta}\right)\over
E_\alpha\left({1\over{\Lambda}^{(i)}_\emptyset}\right)}={{E}_\beta\left({1\over\Lambda^{(i)}_\alpha}\right)\over
{E}_\beta\left({1\over\Lambda^{(i)}_\emptyset}\right)}\nn
\ee
where $\Lambda_\alpha^{(i)}$ is an eigenvalue of the Cherednik Hamiltonian $C_i$ on the non-symmetric Macdonald polynomial $E_\alpha$.
\item The sum of the BA functions over the Weyl chambers $\sum_{v\in S_N}\Psi_m^{(w,v)}(\vec x;\vec y)$ is an eigenfunction of the Cherednik Hamiltonians.
\end{itemize}

\subsection{Non-symmetric versus symmetric triad}

The case of symmetric triad \cite{MMP3} is rather parallel to the non-symmetric case: in that case, there are basically also multiple triads. However, in practice, one needs only one triad, which is the notorious Noumi-Shiraishi power series \cite{NS}. The reason is as follows.

First of all, in the non-symmetric case, the multiple branches are due to essential differences between different $x_i$'s. At the level of reduction to the non-symmetric Macdonald polynomials it results, say, at $N=2$ into additional powers of $x_2$ as compared with $x_1$ at $w=[1,2]$, but not at $w=[2,1]$ (see (\ref{E21}) and (\ref{E12}). At the same time, as soon as the symmetric Macdonald polynomials are symmetric w.r.t. permutations of $x_i$'s, all branches coincide.

As for the different Weyl chambers, indeed, one can consider differently ordered $x_i$ in the Noumi-Shiraishi (NS) function, obtain distinct power series, and associate them with $N!$ distinct Weyl chambers. However, the Ruijsenaars Hamiltonians that define the NS functions (\ref{RS}) are {\bf local}, i.e. do not involve permuted $x_i$'s, and they are {\bf symmetric} w.r.t. any permutation of $x_i$'s.
This results in two facts: (i) the single Noumi-Shiraishi function for any concrete branch alone (without summation) is an eigenvalue of  the Ruijsenaars Hamiltonians; (ii) NS functions associated with distinct Weyl chambers are obtained just my different ordering of $x_i$'s, the coefficients in the power series remain the same for any Weyl chamber.

Because of these reasons, there is effectively just one symmetric triad, which is simultaneously the universal solution, while, in the non-symmetric case, there are many non-symmetric triads, and only sums of them over Weyl chambers form the universal solutions.

\section{Conclusion}

In this paper, we constructed the non-symmetric triad. Similarly to the symmetric triad, it is a power series that has two polynomial reductions: one to the non-symmetric Macdonald polynomials (hence the name), the other one to the (quasi)polynomial BA functions. However, in contrast to the symmetric case, there are $N!$ branches of the non-symmetric triad, and every branch is also parameterized by the choice of the Weyl chamber. In the symmetric case, any choice of the Weyl chamber gave rise to an eigenfunctions of the Ruijsenaars-Schneider Hamiltonians, while only the sum over all Weyl chambers in the non-symmetric case leads to an eigenfunction of the Cherednik Hamiltonians (the universal solution).

Out of $(N!)^2$ triads, $N!(N-1)!$ are independent, since only $(N-1)!$ branches are independent: this is the number of cyclic orbits in the symmetric group $S_N$. In fact, the remaining triads are also not quite independent because of formula (\ref{KS2}).

In this paper, we considered the very explicit and detailed formulas for the power series only for the $N=2$ case. The case of $N=3$ was partly considered in \cite{NSM5}. The power series explicitly in the generic case will be reported elsewhere.

There are two important cases that have been out of consideration so far. First of all, there is a natural extension of the Cherednik integrable systems to the twisted case, at least, at $t=q^{-m}$, $m\in\mathbb{Z}_{\ge 0}$ \cite{NSM1,NSM2,NSM3,NSM4}. However, generalization of the triad to the twisted non-symmetric eigenfunctions remains for the future studies. Note that even the notion of twisted triad in the symmetric case remains undeveloped, though the twisted BA functions are well-defined \cite{ChE}. These twisted BA functions give rise to eigenfunctions \cite{CF,MMP1} of the twisted Hamiltonians associated with various integrable systems in quantum toroidal algebras \cite{MMP}. Moreover, there are also elliptic extensions of the non-twisted symmetric case \cite{MMPZ1,MMPZ2}, however, their extension to non-symmetric case is also unknown.

The second interesting issue is understanding how the triad can be extended to the Cherednik DAHA integrable systems associated with other root systems. There are some problems there with constructing explicit formulas for universal solutions, as was pointed out in \cite{BMP}, and it is unclear to what extend explicit formulas for the triad could be obtained in this case.

At last, there is an interesting issue of mysterious absence of domain walls between different Weyl chambers. This point also deserves further study.

\section*{Acknowledgements}

This work is supported by the RSF grant 26-12-00191.

\end{document}